# OPTIMIZATION OF CRICKET-INSPIRED, BIOMIMETIC ARTIFICIAL HAIR SENSORS FOR FLOW SENSING

*N. Izadi, R. K. Jaganatharaja, J. Floris and G. Krijnen*
Transducers Science & Technology group, MESA+ Research Institute
University of Twente, P.O. Box 217, 7500 AE Enschede, the Netherlands

## ABSTRACT

High density arrays of artificial hair sensors, biomimicking the extremely sensitive mechanoreceptive filiform hairs found on cerci of crickets have been fabricated successfully. We assess the sensitivity of these artificial sensors and present a scheme for further optimization addressing the deteriorating effects of stress in the structures. We show that, by removing a portion of chromium electrodes close to the torsional beams, the upward lift at the edges of the membrane due to the stress, will decrease hence increase the sensitivity.

## 1. INTRODUCTION

Filliform hair based mechanoreceptors on the abdominal appendages (cerci) of crickets are extremely sensitive flow sensors for particle displacement detection with sensitivities bordering thermal noise threshold [1] (Figure 1). The large number of hairs as well as their directivity results in a system capable of complex flow pattern detection, hence recognition of nearby movements with high directional resolution [2]. They make crickets capable of effectively detecting an approaching predator and reacting accordingly. Flow-sensors based on drag-force induced rotation are interesting engineering objects since they allow for high-density arrays which are impractical or impossible with the more common hot wire anemometers (HWA) or microflowns [7].

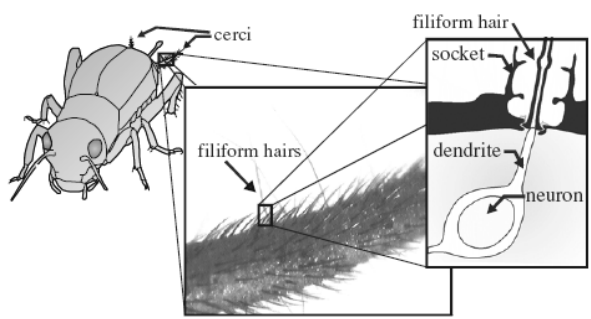

**Figure 1.** Mechanoreceptor hairs on the abdominal appendages of crickets

In recent years mimicking nature to realize more sensitive, robust and reliable sensors has become an appealing practice in micromachining technology. Biomimetic hair structures [3-5] promise low cost, small, sensitive and fast response flow sensors to use in robotic, vehicle control and pathfinder applications, to mention a few.

In previous work we have shown artificial SU-8 hair sensors based on capacitive readout of flow-induced tilting of a silicon nitride membrane with chromium electrodes deposited on top [6]. Figure 2 shows a schematic of these sensors. Fluid flow exerts a drag force on protruding hairs. The resulting drag torque induces a tilting of the hairs from their normal upright position which results in membrane deflection and capacitance changes accordingly.

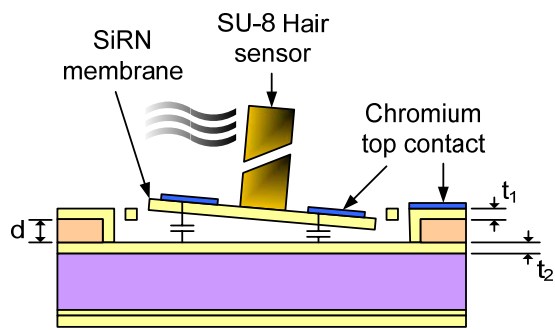

**Figure 2.** Schematic diagram of hair sensors

Figure 4 shows an SEM micrograph of (the base of) a fabricated sensor. It can be clearly seen that the membranes are curves. This leads to a reduced performance.

In this paper we first introduce our capacitive hair sensors' operation principle and determine their sensitivity and electrostatic spring softening coefficient. Then we consider the effect of membrane curvature on the sensitivity and investigate the effects which result in the curvature. To reduce the detrimental effects of curvature we propose a new design for the electrodes and explain the increase in sensitivity accordingly.

## 2. SENSOR MODEL

The mechanoreceptive cercal hair of the crickets has been thoroughly analyzed and its operational mechanics is very often compared to that of an inverted pendulum model [2]. The hair as an inverted pendulum, a typical second-order mechanical system, is described by its mechanical parameters: (i) moment of Inertia $J$, (ii) spring constant $S_0$, and (iii) torsional resistance $R$ (see figure 3).

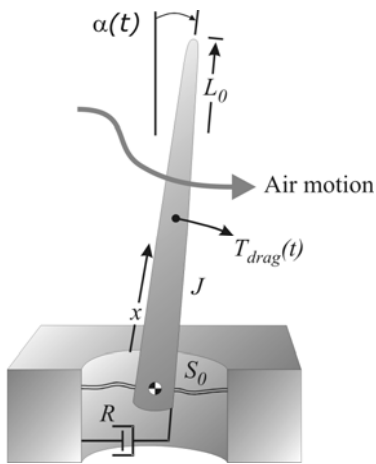

**Figure 3.** Hairs can be modeled as an inverted pendulum which represents a second order mechanical system [1]

As explained before, the recently presented artificial hair sensors [6] are based on the capacitive read-out of flow-induced tilting ($\alpha$) of an SU-8 hair mounted on a silicon nitride membrane. Thin chromium electrodes, deposited on top of the silicon nitride membrane, and the conductive silicon substrate form the top and bottom electrodes of the sensing capacitor, respectively.

The sensitivity of the device is defined as the change in sensor capacitance ($C$) on one half of the plate per unit of tilt-angle ($\alpha$). In the parallel plate approximation it can be analytically derived for small tilt-angles. For a circular membrane it is given for each individual electrode as:

$$\frac{\partial C}{\partial \alpha} = \lim_{\alpha \to 0} \frac{\partial}{\partial \alpha}\left(\iint_A \frac{\varepsilon_0}{d'+\alpha r \sin\varphi} r d\varphi dr\right) \quad (1)$$
$$= \varepsilon_0 \int_0^\pi \int_0^R \frac{-r^2 \sin\varphi}{d'^2} r d\varphi dr = \frac{2}{3} \cdot \frac{\pi \varepsilon_0 R^3}{d'^2}$$

in which

$$d' = d + \frac{t_1 + t_2}{\varepsilon_r} \quad (2)$$

is the dielectric thickness and where the R is the radius of the membrane. In this case R is 85 µm, the gap between the electrodes ($d$) is typically 1 µm, the thickness of the silicon nitride layers $t_1$ and $t_2$ are 1 µm and 0.1 µm respectively, $\varepsilon_0$ and $\varepsilon_r$ are dielectric constant of air and the relative dielectric constant of silicon nitride, respectively.

Additionally, by making use of the electrostatic spring softening effect based on the theory of transduction, the torsional stiffness can be varied on application of a DC bias voltage ($U$) across the electrodes. Then the effective torsional spring stiffness ($S_{eff}$) is given as:

$$S_{eff} = S_0 - U^2 \kappa \quad (3)$$

where, $S_0$ is the torsional stiffness of the springs, without any applied bias voltage and $\kappa$ is the coefficient of effective spring softening given by (applying the same assumptions as give above):

$$\kappa = \frac{1}{2}\frac{\partial^2 C}{\partial^2 \alpha} = \frac{\varepsilon \pi R^4}{2d'^3} \quad (4)$$

By applying DC-bias voltages, the sensor can be adaptively tuned for higher sensitivities. Moreover, by altering the effective torsional stiffness of the springs, the resonance frequency of the structure will be modified, allowing the sensor to adapt to a desired working range. The shift in resonance frequency is given as:

$$\omega_{eff} = \sqrt{\frac{S_{eff}}{J}} = \sqrt{\frac{S_0\left(1-\frac{\kappa}{S_0}\cdot U^2\right)}{J}} = \omega_0 \sqrt{1-\frac{\kappa}{S_0}\cdot U^2} \quad (5)$$

High sensitivity and adaptability of the sensors, which are highly sought, depend mainly on the shape of the sensor capacitance, which in turn relies strongly on the quality of the fabricated membrane and the electrode design. Slight degradation in the quality, for instance, curvature in the silicon nitride membrane, can affect the sensitivity and coefficient of effective spring softening in a deteriorating way.

## 3. STRESS-INDUCED MEMBRANE CURVATURE

The fabrication of the presented hair sensors is discussed in detail in [6]. Large arrays of hair sensors with two primary electrode designs on circular and rectangular membranes were fabricated.

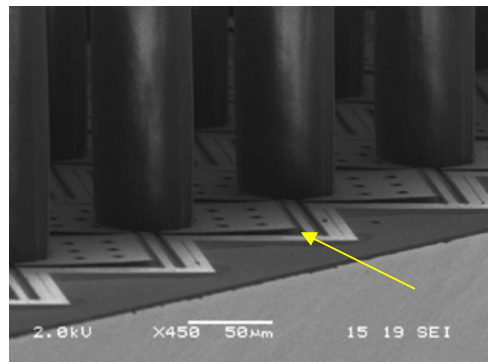

**Figure 4.** SEM image of rectangular membranes. The membranes are curved upward at the edge.

Figure 4 shows the SEM image of rectangular membranes, with a slight curvature. This curvature of the membrane is primarily due to the tensile stress present in the chromium thin layer, deposited on top of the nitride membrane. The curvature results in an undesired increase in the gap between the electrodes, leading to a significant decrease in the sensitivity.

In order to characterize the membrane curvature, the devices were analyzed using a White Light Interferometer (Polytec MSA400) and figure 5 shows the corresponding results. An increase of approximately 3 µm to the intended gap of 1µm, at the curved edges of the membranes is clearly observed.

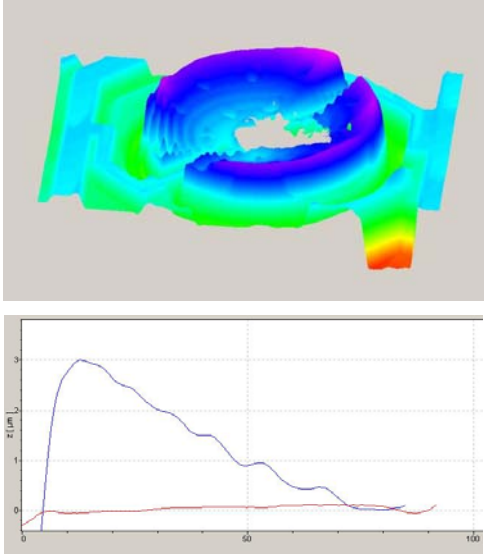

**Figure 5.** White Light Interferometer image from a circular membrane.

The curvature due to the tensile stress depends on the thickness of the chromium layer and increases along the distance from the centre (hair base) to the edges of the membrane.

The curvature can be approximated by a part of a circle with a radius of curvature, $R_c$, which is determined from the measurements. The additional gap due to the stress is:

$$a(r) = R_C \left(1 - \sqrt{1 - \left(\frac{r}{R_C}\right)^2}\right) \approx \frac{r^2}{2R_c} \quad (6)$$

where the last expression only holds for $r \ll R_c$. The curvature leads to an undesirable reduction in sensor capacitance, sensitivity and electrostatic softening coefficient. For example the expression for the sensitivity becomes:

$$\frac{\partial C}{\partial \alpha} = \int_0^R \int_0^\pi \frac{\varepsilon_0 r^2 \sin(\varphi) dr d\varphi}{(d' + \alpha r \sin(\varphi) + a(r))^2} \quad (7)$$

Numerical evaluation of equation (7) shows a great reduction in sensitivity in comparison with equation (1).

## 4. OPTIMIZATION

As we showed in previous section membrane curvature has a great detrimental effect on capacitance, sensitivity and electrostatic spring softening coefficient. For example, the sensitivity of a circular membrane falls off 4 times and the ESS coefficient decreases by 9.5 times due to membrane curvature. Therefore, it is preferable to reduce this curvature. The silicon nitride layer is deposited by an optimized low-stress LPCVD process [8] and hence, the tensile stress is negligible. Therefore it can be considered as a flat substrate. The force due to tensile stress in the chromium layer increases with the thickness. Reducing the electrode thickness is one way to reduce curvature but this has been minimized in the current sensors to 50 nm and there are serious issues regarding electrode resistance on further reduction.

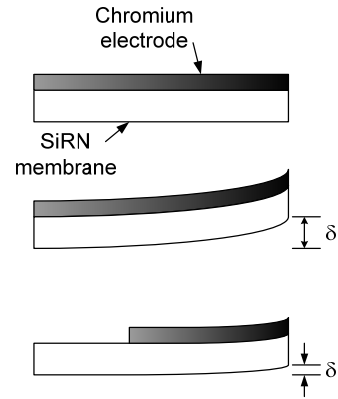

**Figure 6.** Decreasing the electrode length will reduce the upward lift at the edge of membrane.

The electrode area near the torsional beams has minor effect on the sensor's sensitivity since there will be little change in the height of this area on membrane tilt. Therefore, whereas this area plays a major role in the total capacitance, it hardly contributes to the change of capacitance on tilt. On the other hand, by reducing the electrode area as shown in figure 6, the maximum upward lift at the edge of the membrane decreases thereby increasing the sensitivity. Hence it is logical to remove a portion of the electrode close to the torsional beams.

To demonstrate the effectiveness of this design, suppose that the sensitivity has been given by equation (7). To consider the reduced area we change the lower integration limit in these equations to consider an electrode-less area on the membrane which leads to

$$\frac{\partial C}{\partial \alpha} = \int_{r=r_0}^R \int_0^\pi \frac{\varepsilon_0 r^2 \sin(\varphi) dr d\varphi}{(d' + \alpha r \sin(\varphi) + a(r - r_0))^2} \quad (8)$$

We numerically evaluated (9) to investigate the influence of reduced electrode area on the sensitivity.

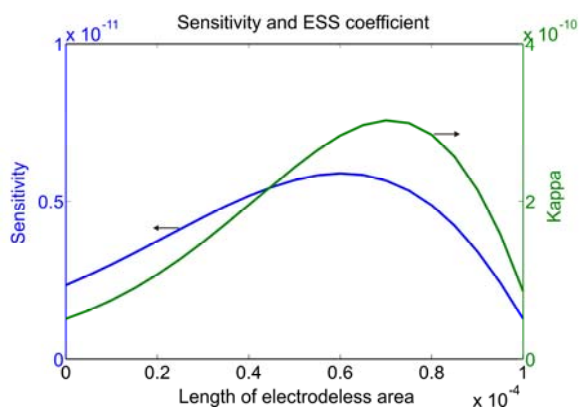

**Figure 7.** Circular membrane sensitivity and ESS changes with respect to length of electrode-less area

Figure 7 shows the actual sensitivity and electrostatic spring softening (ESS) coefficient as a function of the length of the electrode-less area for a circular membrane. Figure 8 shows the same for rectangular membranes.

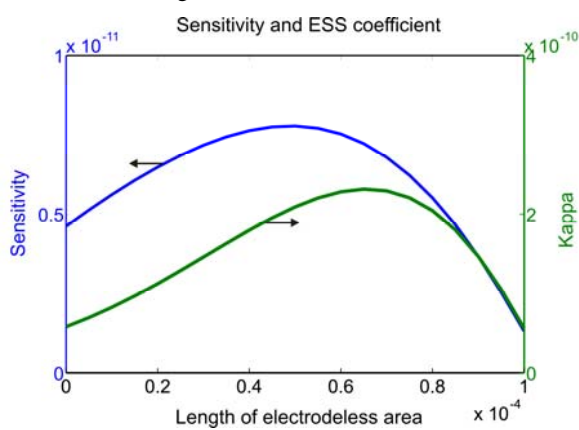

**Figure 8.** Rectangular membrane sensitivity and ESS coefficient change with respect to length of electrode-less area

As it can be seen in the figures, decreasing the electrode length increases the sensitivity initially due to reduction in curvature and reaches a maximum as we remove 60 μm of electrode. Further reduction will decrease the sensitivity as the reduction in the total capacitance becomes a dominant factor. Like the sensitivity, larger kappa is achieved by reducing the area to some extend but again this is not a monotonic increasing function and the maximum is at 70 μm for circular membrane and at 65 μm for rectangular one. The electrode-less area can be optimized to have sensors with desirable sensitivity and electrostatic spring softening coefficient. For a circular membrane, this leads to an improved sensitivity and ESS coefficient by 2.5 and 5.9 times respectively. The table below gives the values of the above discussed parameters.

| Membrane type | Sensitivity | ESS coefficient |
|---|---|---|
| Flat (intended) | $9.50\times10^{-12}$ | $4.89\times10^{-10}$ |
| Curved | $2.35\times10^{-12}$ | $0.51\times10^{-10}$ |
| Reduced electrode | $5.89\times10^{-12}$ | $3.02\times10^{-10}$ |

## 5. CONCLUSION

Stress induced curvature of the electrodes in capacitive biomimicking hairs has a detrimental effect on sensitivity and ESS coefficient. We showed that it is possible to increase both factors by removing a portion of electrodes close to the torsional beams. Upon following this method, there exists a trade off between the capacitance and the sensitivity.

## 6. ACKNOWLEDGEMENTS

The authors want to thank: Meint de Boer and Erwin Berenschot for their advice on processing, Dominique Altpeter for SU-8 processing, Marcel Dijkstra for generating SEM pictures and our colleagues in the EU project CILIA for stimulating discussions and input to this work. The **C**ustomized **I**ntelligent **L**ife-**I**nspired **A**rrays project is funded by the Future and Emergent Technologies arm of the IST Programme.